\documentclass[10pt,conference]{IEEEtran}

\usepackage{listings}
\usepackage[inline]{enumitem}
\usepackage{xspace}
\usepackage{lscape}
\usepackage{multirow}
\usepackage{lipsum}
\usepackage{cite}
\usepackage{xcolor}
\usepackage{url}
\usepackage{graphicx}

\newcommand{\fsl}{\textsl}
\newcommand{\ftt}{\texttt}

\newcommand{\etal}{{et al.\@\xspace}}

\setlist[description]{leftmargin=1em}
\setlist[itemize]{leftmargin=1em}
\setlist[enumerate]{leftmargin=1.5em}
 
\definecolor{codegreen}{rgb}{0,0.6,0}
\definecolor{codegray}{rgb}{0.5,0.5,0.5}
\definecolor{codepurple}{rgb}{0.58,0,0.82}
\definecolor{backcolour}{rgb}{0.95,0.95,0.92}

\lstdefinelanguage{JavaScript}{
  keywords={break, case, catch, continue, debugger, default, delete, do, else, false, finally, for, function, if, in, instanceof, new, null, return, switch, this, throw, true, try, typeof, var, void, while, with},
  morecomment=[l]{//},
  morecomment=[s]{/*}{*/},
  morestring=[b]',
  morestring=[b]",
  ndkeywords={class, export, boolean, throw, implements, import, this},
  keywordstyle=\color{blue}\bfseries,
  ndkeywordstyle=\color{darkgray}\bfseries,
  identifierstyle=\color{black},
  commentstyle=\color{purple}\ttfamily,
  stringstyle=\color{red}\ttfamily,
  sensitive=true
}

\begin{document}

\title{An Extended Model of Software Configuration} 


\author{\IEEEauthorblockN{Rezvan Mahdavi-Hezaveh}
\IEEEauthorblockA{North Carolina State University\\
Raleigh, NC, USA\\
Email: rmahdav@ncsu.edu}
\and
\IEEEauthorblockN{Sameeha Fatima}
\IEEEauthorblockA{North Carolina State University\\
Raleigh, NC, USA\\
Email: sfatima3@ncsu.edu}
\and
\IEEEauthorblockN{Laurie Williams}
\IEEEauthorblockA{North Carolina State University\\
Raleigh, NC, USA\\
Email: laurie\_williams@ncsu.edu}}

\maketitle

\begin{abstract}

Both feature toggles and configuration options are modern programmatic techniques to easily include or exclude functionality in a software product. The research contributions to these two techniques have most often been focused on either one of the techniques separately. However, focusing on the similarities of these two techniques may enable a more fruitful combined family of research on \textit{software configuration}, a term we use to encompass both feature toggles and configuration options. 
Additionally, a common terminology may have enabled meta-analysis, a more practical application of the research on feature toggles and configuration options, and prevented duplication of research effort. \emph{The goal of this research study is to aid researchers in conducting a family of research on software configuration by extending an existing model of software configuration that provides terminology and context for research studies.} To achieve our goal, we started with Seigmund et al.'s Model of Software Configuration (MSC) which was developed based on interviews and publications on configuration options. We explicitly extend the MSC to include feature toggles and to add 
qualitative analysis (using deductive and inductive coding) of feature toggle-related software repositories and publications.
From our analysis, we propose MSCv2 as an extended version of MSC. We evaluated MSCv2 through its application on five academic publications and the Chrome industrial system. Our results indicate that multiple researchers studying the same software system may provide different definitions of software configuration in their publications. Also, similar research questions may be answered on feature toggles and on configuration options repeatedly because of a lack of a clear definition of software configuration in research publications. These observations indicate that having a model for defining software configuration may enable more clear, comparable, and generalized research on the software configuration family of research. Practitioners may also benefit when defining software configuration in their systems to better knowledge transfer to other practitioners and researchers.
\end{abstract}


\IEEEpeerreviewmaketitle

\section{Introduction}
\label{introduction}

Both feature toggles and configuration options are techniques that practitioners use in the modern software development process that have similar definitions. \textit{Feature toggling} involves checking the value of a variable in a conditional statement to change the behavior of the system without changing the code. Feature toggles are used in continuous integration/continuous delivery (CI/CD) pipelines, gradually rolling out software updates and running experiments on new features~\cite{mahdavi2021software}. 
\textit{Configuration options} are key-value pairs that allow software customization by including or excluding functionality in a software system~\cite{meinicke2019exploring}. Using configuration options allows developers to implement software product lines~\cite{siegmund2020dimensions,meinicke2019exploring, apel3feature}. The border between feature toggle and configuration option concepts is blurred. Based upon an interview study, Meinicke \etal~\cite{meinicke2019exploring} found that practitioners may both consider feature toggles as a use of configuration options for a new purpose, such as CI/CD; or consider configuration options as a subset of feature toggles.  
Due to difficulties to separate the two concepts, both can be considered \textit{software configuration}~\cite{meinicke2019exploring,siegmund2020dimensions}.

Researchers perform studies in both feature toggles and configuration options research areas. Based on semi-structured interviews, Meinicke~\etal~~\cite{meinicke2019exploring}found that using these two techniques may cause similar problems that have similar solutions. So, if a research question is answered in the configuration option area, the result may also be applicable to feature toggles and vice versa. 
\textit{Ignoring similarities and differences between the characteristics of software configuration in research studies can lead to duplication of effort, inefficiency in research, and unclear comparisons and generalizations.}

A family of research can benefit from a framework or common ontology to enable meta-analysis. Research shows the necessity of having a common ontology to exchange and organize knowledge
~\cite{layman2004exploring, nour2003ontology}. Researchers have proposed evaluation frameworks in software engineering scopes, such as Extreme Programming (XP)~\cite{williams2004toward, williams2004extreme} and software security~\cite{morrison2017security} to enable comparing case studies and results in related publications. 
Runeson~\etal~\cite{runeson2012case} explains that using a theoretical framework to design case study research in software engineering makes the context of the research clear, helps researchers to conduct the study and build upon the results. 
Hence, software configuration researchers can benefit from a common ontology. While prior work~\cite{williams2004toward, williams2004extreme, morrison2017security, runeson2012case} has used the terms \textit{framework} or \textit{ontology}, because we are adapting the \textit{Model of Software Configuration}~\cite{siegmund2020dimensions}, henceforth we utilize the term \textit{model} to refer to the artifact we are proposing to enable meta-analysis in a family of software configuration research.

Siegmund \etal~\cite{siegmund2020dimensions} derived a Model of Software Configuration (MSC) consisting of eight \textit{dimensions} that are high-order topics related to configurations, and a total of 47 \textit{values} assigned to these dimensions. The MSC was developed by interviews and the analysis of configuration options-related publications. 
In another study, Meinicke {\etal}~\cite{meinicke2019exploring} explored similarities and differences between configuration options and feature toggles by conducting semi-structured interviews with practitioners. They identified 10 \textit{themes} that explain the differences between feature toggles and configuration options. We use these two publications as the basis of our study and we will discuss them in details in Subsections~\ref{sec:themes}, and~\ref{sec:msc}. 

\emph{The goal of this research study is to aid researchers in conducting a family of research on software configuration by extending an existing model of software configuration that provides terminology and context for research studies.} We call the \fsl{extended Model of Software Configuration} MSCv2 in the rest of this paper. We state the following research questions:

\begin{description}
    \item[RQ1:] What dimensions and values need to be added to extend the MSC to MSCv2 to cover feature toggles as well as configuration options?
    \item[RQ2:] Can the MSCv2 be used to model software configuration in research publications and in industrial systems?
\end{description}

To answer RQ1 and extend the MSC~\cite{siegmund2020dimensions}, we qualitatively (using deductive and inductive coding) analyzed the documentation of 20 feature toggle management systems 
repositories on GitHub, and the 
8 feature toggle-related publications. 
While analyzing these resources, we recorded new values and new dimension. Our analysis of the feature toggle-related resources 
triangulates~\cite{runeson2012case} on the dimensions and values of the MSC and adds additions dimension and values which enable the model to be applicable to both feature toggles and configuration options.
 As a result of answering RQ1, we extend MSC and developed MSCv2 with one new dimension and 24 new values.
To answer RQ2, we selected two publications on 
Chrome~\cite{rahman2016feature, rahman2018modular}, three publications with similar research questions~\cite{sayagh2018software,mahdavi2021software,rahman2016feature}), and the Chrome repository\cite{chromiumrepo} as an industrial system. We applied MSCv2 to these artifacts by creating instances of MSCv2 and populating the dimensions and values, as we will explain in Section in~\ref{method_phase3}. We discuss our observations on the applicability of MSCv2 in Section~\ref{discussion}.

The contributions of this paper are as follows:
\begin{enumerate}
    \item An extended Model of Software Configuration (MSCv2) with one new dimension and 24 new values which is applicable to both feature toggles and configuration options;
    \item Software configuration models in related publications;
    \item Software configuration models of the Chrome repository;
    \item A list of feature toggle management systems on GitHub.
\end{enumerate}

The rest of the paper is organized as follows: in Section~\ref{background}, we describe the background of the research area and prior related work. In Section~\ref{method}, we explain our methodology. In Section~\ref{model}, we report our findings on MSCv2. In Section~\ref{discussion}, we explain the application of MSCv2. We enumerate the limitations in Section~\ref{limitations}. We conclude the study in Section~\ref{conclusion}.

\section{Background and related work}
\label{background}

This section describes the background and related work.

\subsection{Background}
\label{background-background}
Using feature toggles is a technique used by companies practicing Continuous Integration (CI) and Continuous Delivery (CD)~\cite{parnin2017top,rahman2015synthesizing}.
Programming languages have provided the constructs to implement feature toggles for a long time. However, the first documented use of feature toggles to support CI/CD was at Flickr in 2009 ~\cite{rossflippingout}. Listing~\ref{lst:toggle} is an example of a feature toggle. The value of the toggle \ftt{useNewAlgorithm} is checked to determine which search algorithm to call\cite{mahdavi2021software}. 

Configuration options are key-value pairs to include functionality in a software system or setting parameters such as buffer size~\cite{meinicke2019exploring}.
 Based on the definition of configuration options and feature toggles by researchers~\cite{meinicke2019exploring, rahman2016feature, hodgsonfeaturetoggle}, configuration options can be considered as a subset of the feature toggles or feature toggles can be seen as configuration options that are used for new purposes. 
Therefore, the concepts of configuration options and feature toggles have similar definitions and have similar challenges. 
Hence, researchers can benefit from a model of software configuration that covers both concepts to prevent duplication of effort and generalize research results.

\begin{lstlisting}[caption={An example of a feature toggle ~\cite{mahdavi2021software}.}, label={lst:toggle}]
function Search() {
    var useNewAlgorithm = false;
    if(useNewAlgorithm) {
      return newSearchAlgorithm(); }
    else {
      return oldSearchAlgorithm(); } }
\end{lstlisting}


\subsection{Related Work}
\label{background-relatedworks}

In this section, we explain related work.
~We explain Meinicke {\etal}~\cite{meinicke2019exploring} in Section~\ref{sec:themes}, and Siegmund {\etal}~\cite{siegmund2020dimensions} in Section~~\ref{sec:msc} as the foundational basis of our study.

\subsubsection{Feature toggles}
\label{relatedworks-toggles}


Rahman {\etal}~\cite{rahman2016feature} performed a thematic analysis of gray literature artifacts provided by practitioners to understand the challenges, and advantages of using feature toggles in practice. In addition, they quantitatively analyzed the usage of feature toggles in 39 releases of Chrome over 5 years.
Rahman {\etal} reported three purposes of using feature toggles: rapid release,  trunk-based development, and A/B testing. They found that using feature toggles 
can introduce technical debt and require additional maintenance effort.

Rahman {\etal}~\cite{rahman2018modular} extracted four architectural representations of Chrome: conceptual, concrete, browser reference, and  feature toggle architecture. Their results indicate that developers can use the extracted feature toggle architecture to determine which feature toggle affects which module and which module is affected by which feature toggle.

Mahdavi-Hezaveh {\etal}~\cite{mahdavi2021software} analyzed 109 grey literature artifacts and publications on feature toggles and identified 17 feature toggle usage practices in four categories: Management, Initialization, Implementation, and Clean-up. They conducted a survey with feature toggle expert practitioners to validate their identified practices. They observed that using a management system to manage feature toggles is a popular practice in industry. Though practitioners are aware of the importance of removing the toggles after the purpose of using it is done, the least used practices belong to the clean-up category. 

Meinicke {\etal}~\cite{meinicke2020capture} proposed a semi-automated approach to detect feature toggles in open-source repositories by analyzing the repositories' commit messages. They provided a dataset of 100 GitHub repositories that use feature toggles.
To show the benefits of using their dataset, they analyzed short-lived toggles in repositories and found the importance of having a toggle owner to shorten the lifetime of a feature toggle.

Ramanathan {\etal}~\cite{ramanathanpiranha} developed an automated code refactoring tool to delete code of unused feature toggles in Uber repositories. Their tool analyzes the abstract syntax tree of the code, generates a diff on the GitHub repository, and assigns the diff to the developer who creates the feature toggle
Over 1.5 years, the tool generated diffs for 1381 feature toggles, and 65\% of them were accepted by developers with no changes.

Hoyos~{\etal}~\cite{hoyos2021removal} analyzed the source code of 12 Python repositories that use feature toggles and surveyed 61 practitioners to shed light on the removal of feature toggles in practice. They found that 75\% of feature toggles in are removed within 49 weeks after creation. They also found that not all feature toggles are removed from code and some long-lived toggles are not designed to live that long. In the survey, practitioners reported removing feature toggles in situations such as when the feature stabilized in the production.

Prutchi~\etal~\cite{prutchi2022adoption} studied the effect of adopting feature toggles on the frequency and complexity of branch merges, the number of defects, and their fix time in 949 open-source repositories. They observed a decrease in the merge effort and an increase in defect fix time. However, they did not confirm this increase is because of adopting feature toggles.

Mahdavi-Hezaveh~\etal~\cite{mahdavi2022feature} conducted a qualitative analysis of 60 GitHub repositories, proposed 7 heuristics for structuring feature toggles in the code, and identified 12 metrics to support the heuristics. They conducted a case study on 80 repositories to analyze relations between heuristics and metrics. Their results showed that practitioners should have self-descriptive toggles, 
and ensure removal of toggles.

\subsubsection{Configuration options}
\label{relatedworks-options}

Sayagh {\etal}~\cite{sayagh2018software} interviewed 14 practitioners, surveyed software engineers, and did a systematic literature review on configuration options' publications to understand process of managing configuration options. They identified 9 activities, 22 challenges
, and 24 best practices.

Meinicke {\etal}~\cite{meinicke2016essential} analyzed 8 small- to medium-sized configurable systems' traces to identify interactions of configuration options. 
They showed that configuration interactions is lower than combinatorial complexity in theory.

Zhang {\etal}~\cite{zhang2021evolutionary} studied 1178 configuration-related commits of four open-source cloud system repositories over a period of 2.5 years. They studied code changes to help reduce misconfigurations and developed a taxonomy for configuration design and implementation evolution in cloud systems.



In these related work, although the researchers define feature toggle or configuration option, the definition of software configuration in their research could be more explicit by the use of a model as a common language for defining software configuration. In this paper, we extend and apply such a model.

\section{Methodology}
\label{method}

\begin{figure}[!htb]
\centering
\includegraphics[width=0.9\columnwidth]{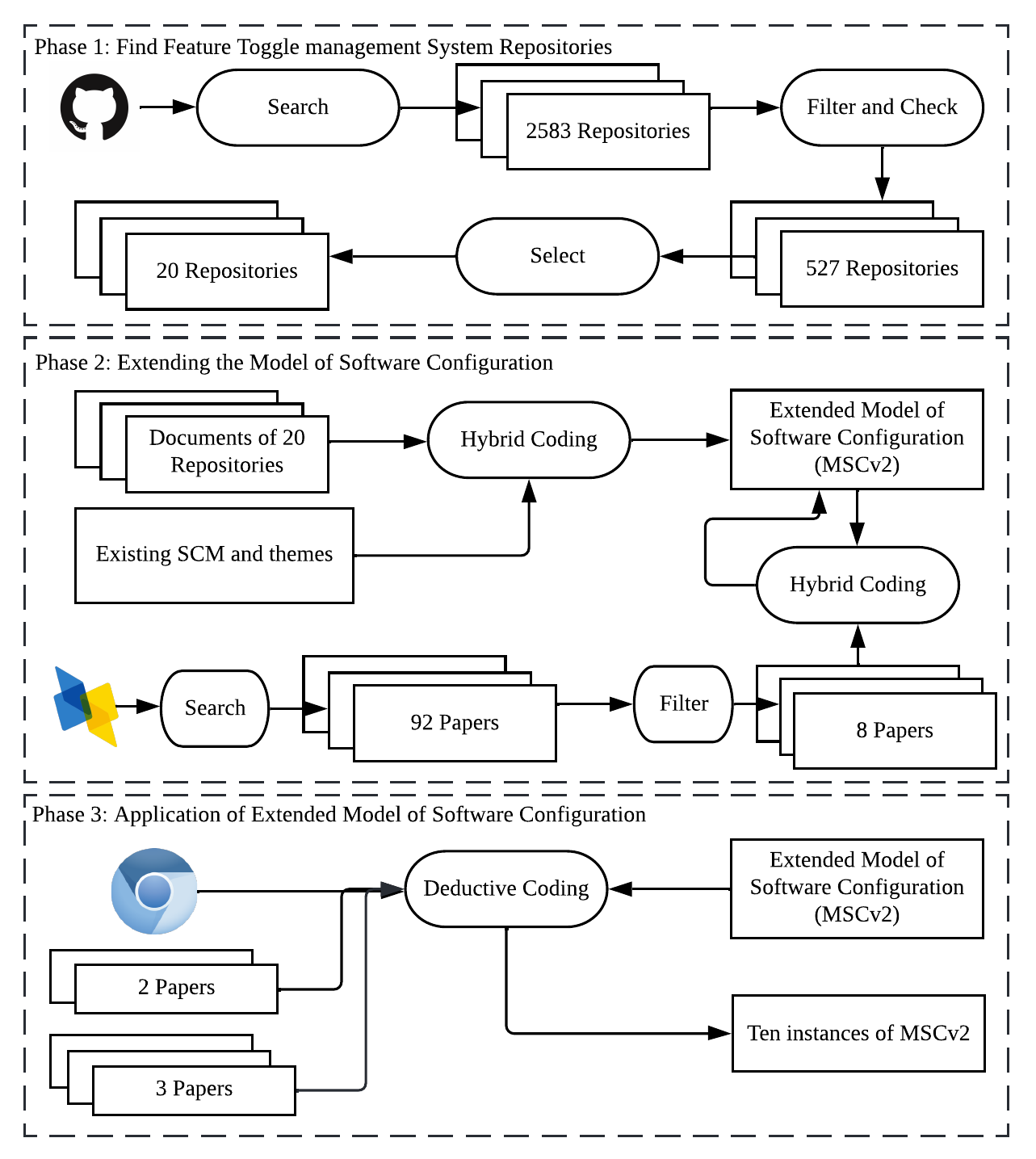}
\caption{Research methodology outline.}
\label{fig:methodology}
\end{figure}

In Sections~\ref{sec:themes} and \ref{sec:msc}, we explain the two foundational studies.  We explain three phases of our methodology in Sections~\ref{method:one}--~\ref{method_phase3}.  Figure~\ref{fig:methodology} outlines this methodology.

\subsection{Meinicke {\etal}:  Feature Toggles vs. Configuration Options}
\label{sec:themes}

To better understand feature toggles and specify their relation to configuration options, Meinicke {\etal}~\cite{meinicke2019exploring} conducted nine semi-structured interviews with practitioners who were feature toggle experts. By analyzing the interview scripts, they identified ten themes that explain the differences between configuration options and feature toggles. They also reflected that ``depending on one’s perspective, configuration options can be seen as a special subset of feature toggles or one can see feature toggles as using configuration options for a new purpose.'' Hence, we consider feature toggles and configuration options to be part of the same branch of knowledge. We use Meinicke~{\etal}'s themes as a basis for our research study.

\subsection{Siegmund {\etal}: Model of Software Configuration (MSC)}
\label{sec:msc}

To provide software configuration terminology for research studies and help practitioners to find possible challenges, Siegmund {\etal}~\cite{siegmund2020dimensions} derived the MSC that consists of eight dimensions for software configuration with 47 values in total.  \textit{Dimensions} are high-order topics related to configurations and one or more \textit{values} can be assigned to each dimension. They used a mixed-methods approach. First, they interviewed 11 practitioners, such as developers, team leads, and senior software engineers. From the collected interviews' data, they created an initial model of configuration. Their initial model includes seven dimensions with 32 values. Then, they analyzed two related publications to their study to validate their results and complete their model. As a result, they found one additional dimension and 15 new values. To verify the applicability of their developed model and find gaps in the area, they applied their model to 16 \textit{configuration option}-related publications. 

We use MSC as a basis for our research study and extend it explicitly to support both feature toggle and configuration options which we refer to jointly as \textit{software configuration}. 


\subsection{Phase One: Find Feature Toggle Management Systems} 
\label{method:one}
A feature toggle management system is a platform that helps practitioners to define, implement, integrate, and manage feature toggles in their code. Feature toggle management systems often provide services, such as APIs, an admin UI, data storage for toggles, and auditing logs. To find open-source feature toggle management system repositories, on 14th April 2021, we searched GitHub using the search API with the following keywords (feature toggles are also called feature flags, feature switches, and feature flips in literature~\cite{fowlerfeaturetoggle}): ``feature toggle'' (1,065 repositories), ``feature flag'' (1,045 repositories), ``feature switch'' (355 repositories), and ``feature flip'' (118 repositories). 
In total, we retrieved 2,583 repositories. To filter the results, we follow the steps below (the numbers in parenthesis are the number of the repositories after each step): (1) We removed duplicated repositories (2,387), (2) We removed repositories with zero stars (910), (3) We removed repositories with no update in last five years, after 2015-12-31 (765), (4) We read the name and description of repositories and removed repositories that are not about feature toggle management. One example is TriPlayer~\cite{triplayer}. The description of this repository is ``A feature-rich background audio player for Nintendo Switch (requires Atmosphere)''. Because of having words \textit{feature} and \textit{switch} in the description, the repository exists in our retrieved result. However, it is not related to the \textit{feature switch} concept, so we removed it. (548), (5) We removed repositories that were archived by the owners (527). The top five programming languages of the final 527 repositories are JavaScript (110), PHP (62), TypeScript (61), C\# (52), and Ruby (49).


\subsection{Phase Two: Extending the MSC (MSCv2)}
\label{method:two}

In Phase~2, we extend the MSC by qualitative analysis (using deductive and inductive coding) of feature toggle-related resources including the documentation from the feature toggle management system repositories, and 
related publications. 

To select a subset of Phase~1 results for document analysis, we focused on the top 5 programming languages with the greatest number of repositories mentioned in Section~\ref{method:one}. 
We sorted repositories based on their stars. Then we selected the top four ranked repositories for each of the 5 programming languages (20 repositories~\cite{unleash,rollout,flipper,FeatureToggle,fflip,FeatureManagementDotnet,qandidatetoggle,flag,featureflags,flags,react-feature-toggles,flip,phprollout,FeatureSwitcher,flagged,Esquio,swivel,feature,ngx-feature-toggle,FeatureSwitch}) and analyzed their documentation as discussed below. 
We did not find any new values from the fourth-ranked repositories
, so we stopped. 

To extend the MSC, the first and second authors qualitatively analyzed the available resources from each one of the 20 selected repositories including the README files, documentation, videos, and websites. 
We used a hybrid coding approach including \fsl{deductive coding} and \fsl{inductive coding}. Coding is a technique for qualitative analysis of the text by assigning words or short phrases as codes to parts of the text~\cite{saldana2015coding}.
Deductive coding is an approach in which codes are developed ahead to guide the coding process, and inductive coding is an approach in which codes are developed while analyzing the text~\cite{christians1989logic}. 
In our hybrid coding approach: (1) We used Siegmund {\etal}'s dimensions and Meinicke {\etal}'s themes as predefined codes (deductive coding); (2) We assigned the predefined codes to the text of documentation for each repository; (3) At the same time, we assigned new codes to any parts of the text that was not covered by predefined codes (inductive coding); (4) In steps 2 and 3 we recorded values for each code (codes are dimensions or themes, and each one has possible values), (5) We selected the most relevant new codes and values from step 3
, and added them to MSC to create MSCv2. The first and second authors of the paper performed steps 1 to 5 separately, discussed their findings at the end of the analysis, and resolved disagreements in their codes. During this process, the first author found one possible new dimension, and the second author found two possible new dimensions. After discussion, the authors agreed on one of the new dimensions and discard the other two. 
The Cohen’s Kappa~\cite{viera2005understanding} agreement scores on values of existing dimensions are 0.77, 0.58, 0.45, 0.28, 0.43, 0.47, 0.8, 0.33 in order for dimensions listed in Section\ref{model}. All the scores are between ``fair agreement'' and ``substantial agreement''. During this analysis, we found one new dimension and 18 new values.

After analyzing the repositories' documentation, we use MSCv2 to model the software configuration of the analyzed systems in relevant feature toggle publications to evaluate the comprehensiveness of the model and to find new dimensions and values. To find feature toggle-relevant publications, we searched DBLP (Digital Bibliography \& Library Project) on 22nd March 2022 with the following keywords: ``feature toggle'' (9 with 2 duplicates), ``feature flag'' (13), ``feature switch'' (75), and ``feature flip'' (6). The first and second authors read the title of peer-reviewed publications and selected related ones for our study. For any publication with an unclear title, we read the abstract of the publication to find out if it is relevant or not.
From these, eight publications are explained in Section~\ref{relatedworks-toggles} and one publication is one of the basis studies of this paper~\cite{meinicke2019exploring} as explained in Section~\ref{sec:themes}. The first and second authors of the paper, (1) Read each of the eight publications; (2) Performed deductive coding with the list of codes (dimensions) they found in documentation analysis; (3) Recorded any values mentioned for dimensions; and (4) Performed inductive coding for new aspects that was not covered by predefined codes. The first and second authors discussed their results and resolved disagreements. The Cohen’s Kappa agreement scores~\cite{viera2005understanding} for values of dimensions are 0.77, 0.40, 0.22, 0.82, 0.31, 0.33, 0.41, 0.51, and 0.06 in order for dimensions listed in Section~\ref{model}. All the scores are between ``slight agreement'' and ``substantial agreement''. During this process, we found six new values and no new dimensions
Our process showed that the software configuration can be modeled systematically in related publications.

\subsection{Phase Three: Application of MSCv2 
}
\label{method_phase3}
Researchers and practitioners can follow the steps explained in the following subsections~\ref{sec:apply-academia} and~\ref{sec:apply-industry} to apply MSCv2 to their research studies' or industrial systems' design and evaluation. \textit{Application} means creating an instance of MSCv2 with appropriate dimensions and values.

\subsubsection{Academia}
\label{sec:apply-academia}

We analyzed two sets of publications. The first set includes two feature toggle publications on Chrome \cite{rahman2016feature, rahman2018modular}. The second set includes three publications asking similar research questions about practitioners' practices in using software configurations~\cite{rahman2016feature,mahdavi2021software,sayagh2018software}: two publications focused on feature toggles and one on configuration options.

For all five publications, the first and second authors of the paper, (1) Read the publication; (2) Extracted parts of the text explaining characteristics of configuration in their research including the definition of software configuration in the literature and in analyzed industrial system; (3) Performed deductive coding on the extracted text with the list of dimensions and values found in Phase~2; and (4) visualized the result by creating an instance of the MSCv2 with identified values for each dimension. 
After modeling the configuration using MSCv2, we compare the models
, and discuss in Section~\ref{discussion}.

\subsubsection{Industry}
\label{sec:apply-industry}

To show the applicability of MSCv2 in practice, we modeled configurations as they are defined in the Chrome repository. The first and second authors of the paper, (1) Searched \fsl{docs} directory of Chrome repository by using ``feature toggle'', ``feature flag'', ``feature switches'', ``feature flips'', and ``configuration options'' keywords; (2) Found a file explaining five terms for configurations in the repository~\cite{chromium}
; (3) Performed deductive coding on the definitions on this file with the list of dimensions and values found in Phase~2, and identify values for each dimension for each term; (4) Follow links in the file to other files, and found more documentation; (5) Performed steps 3 and 4 until no new value found for dimensions; and (6) visualized the five models by creating an instance of the MSCv2 with identified values for each dimension.
We discuss our observations in Section~\ref{discussion}.

\section{The Extended Model of Software Configuration (MSCv2)}
\label{model}

\begin{figure}[!htb]
\centering
\includegraphics[width=1.0\columnwidth]{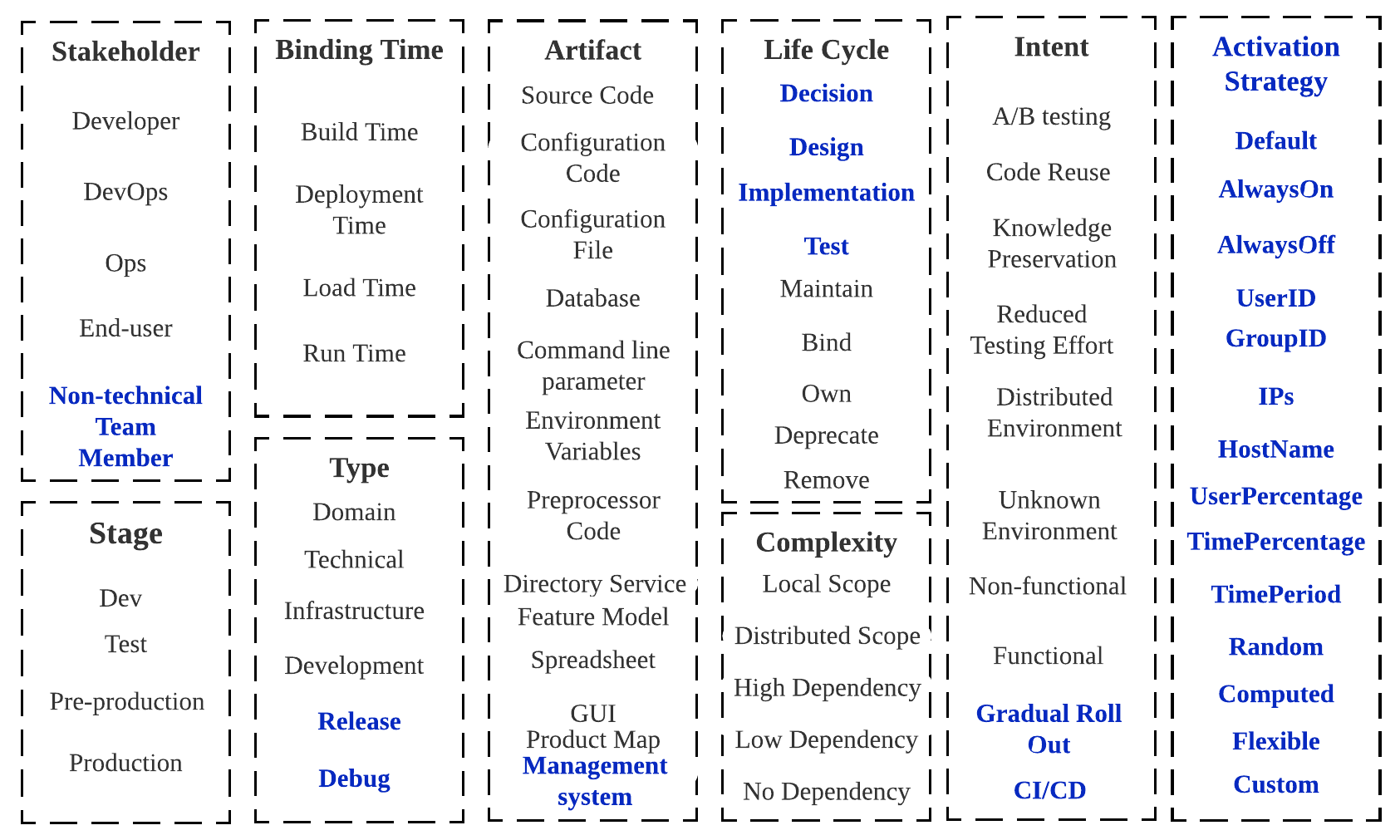}
\caption{MSCv2: Extended Model of Software Configuration} 
\label{fig:extendedmodel}
\end{figure}

In Phase~2 of our methodology, we extended the MSC such that it can be applied to both feature toggle and configuration option concepts. 
Our analysis of feature toggles resources revealed one new dimension (Activation Strategy) with 14 values. We also identified 10 additional values for the existing dimensions. Figure~\ref{fig:extendedmodel} shows the MSCv2 with 9 dimensions and 70 values. The blue (and bold) items are the new dimensions and new values added to the MSC (\textbf{answer to RQ1}). 

In the following subsections, we describe the dimensions in MSCv2.
For each dimension, first we define the dimension (\fsl{Definition}). The definition can come from \cite{siegmund2020dimensions}, \cite{meinicke2019exploring}, or from our analysis. Next, we explain the possible values based on two basic studies and our analysis of documents from feature toggle management systems' repositories (\fsl{Model Extension}). Then, we explain the result of the application of MSCv2 to feature toggle-related publications to find missed dimensions and values (\fsl{Model Completion}). Lastly, we summarize each dimension in a box. 
If the dimension or values came from \cite{siegmund2020dimensions}, we marked them with a S1; from \cite{meinicke2019exploring} with a S2; and from our observations and analysis with a S3 identifiers. 

\subsection{Stakeholder (S1, S2, S3)}

\fsl{Definition.} Based on Siegmund {\etal} \cite{siegmund2020dimensions}, the stakeholder is anyone ``who deals with configuration''. Meinicke {\etal} \cite{meinicke2019exploring} defines the stakeholder as the person ``who is making configuration decisions''. 

\fsl{Model Extension.} In the MSC, \fsl{Developer}, \fsl{DevOps}, \fsl{Ops}, and \fsl{End-user} can change the software configurations of the system based on their needs. However, based on our analysis of repositories' documents, \fsl{End-user} is not a stakeholder in feature toggle usage. We have observed \fsl{Developer}, \fsl{DevOps}, and \fsl{Ops} as stakeholders in the analyzed documentation, but have not observed the possibility for users to change the value of or manage toggles. 
Based upon their study, Meinicke \etal~\cite{meinicke2019exploring} found a key difference between configuration options and feature toggles is the person who makes configuration decisions: feature toggles are generally controlled by \fsl{Developers} and \fsl{Ops}, and configuration options are generally controlled by \fsl{End-users}. However, they found that feature toggles may be exposed to \fsl{End-users} in the use of toggles to conduct experiments. For example, feature toggles are exposed to users in chrome://flags/ where users of Chrome can enable or disable features themselves. As a result, the distinction between feature toggles and configuration options is blurred. 
%
Moreover, while we analyze the documents for stakeholders, we have identified a new value: \fsl{Non-technical Team Member} \cite{flipper}
who is able to release changes while using feature toggles. 

\fsl{Model Completion.} 
We have observed all the values for this dimension in analyzed publications. Related to the new value \fsl{Non-technical Team Members}, Mahdavi-Hezaveh \etal~\cite{mahdavi2021software} identified ``Give access to team members'' practice in which permission is given to Q\&A team, sales team, and product managers to prevent the management bottleneck of toggles~\cite{mahdavi2021software}.
%

\noindent\framebox[\linewidth]{
  \parbox{0.97\linewidth}{
\textbf{Stakeholder (S1, S2, S3)}
%
is a person who is making configuration decisions. Values 
are: Developer (S1, S2, S3), DevOps (S1, S3), Ops (S1, S2, S3), End-user (S1, S2), and Non-technical Team Member (S3).}}

\subsection{Stage (S1, S3)}

\fsl{Definition.} Stage ``describes that configuration happens in different stages of the development process'' \cite{siegmund2020dimensions}.
\\ 
\fsl{Model Extension.} Siegmund {\etal} identified four Stage values: Dev, Test, Pre-Production, Production.

In our analysis of feature toggle management systems, we observed all four stages of feature toggle usage. Using release toggles for dark launches and trunk-based development points to the \fsl{Production} stage. Using toggles for gradual rollout points to \fsl{Pre-production} stage. Ops toggles and permission toggles are used in the \fsl{Production} stage. Development toggles are used in \fsl{Dev}, and experiment toggles are used in \fsl{Test} stage. 

\fsl{Model Completion.} Our publications' analysis confirms our observation in feature toggle management systems. 

\noindent\framebox[\linewidth]{
  \parbox{0.97\linewidth}{
\textbf{Stage (S1, S3)}
describes that configuration happens in what stages of the development process.
Values 
are: Dev (S1, S3), Test (S1, S3), Pre-Production (S1, S3), Production (S1, S3)
}}

\subsection{Binding Time (S1, S3)}

\fsl{Definition.} Binding time is the time of binding a value to a software configuration \cite{siegmund2020dimensions}.

\fsl{Model Extension.} The binding time of software configurations in a system will affect the implementation and design of the configuration. Siegmund {\etal} identified four binding time values: Build Time, Deployment Time, Load Time, and Run Time. For example, \fsl{Build Time} configurations are used in form of build scripts; \fsl{Deployment Time} configurations are used as configuration files; \fsl{Load Time} configurations are used in forms of properties files; and \fsl{Run Time} configurations are used in form of databases, or user interfaces. Based on our analysis of configuration management system documentation, \fsl{Run Time} is the most used (but not the only) binding time for feature toggles. 
In contrast, we have not observed any \fsl{Build time} as the binding time for toggles.

\fsl{Model Completion.} In the analyzed publications, we have the same observation about the binding time of feature toggles. Authors of all publications except one \cite{meinicke2020capture} mentioned \fsl{Run time} as the binding time for feature toggles. 

\noindent\framebox[\linewidth]{
  \parbox{0.97\linewidth}{
\textbf{Binding Time (S1, S3)}
is the time of binding a value to a software configuration. 
Values 
are: Build Time (S1), Deployment Time (S1, S3), Load Time (S1, S3), Run Time (S1, S3).}}

\subsection{Type (S1, S3)}

\fsl{Definition.} Siegmund {\etal}~\cite{siegmund2020dimensions} define type of configuration as ``what should be configured''.  

\fsl{Model Extension.} The values Siegmund {\etal} identified for this dimension are: Domain, Technical, Infrastructure, and Development. \fsl{Domain} configurations are used to change software based on the user's requirements. \fsl{Technical} configurations (that contains \fsl{Infrastructure} and \fsl{Development} types) focus on the environment of the system and its deployment process. \fsl{Infrastructure} configuration is used to adjust software to related hardware and software, such as database systems and used ports. \fsl{Development} configuration refers to setting-up development tools, such as IDEs, build tools, building and testing process, and automating the deployment process. 

In the feature toggle management system documentation, we observed \fsl{Domain} (in all repositories) and \fsl{Infrastructure} (in \cite{flag}), but no mention of \fsl{Development} type. We observed the need to define new values for this dimension to cover some types of feature toggles provided by management systems, such as toggles to enable trunk-based development~\cite{flipper}.

\fsl{Model Completion.} Based on literature in the analyzed publications~\cite{rahman2016feature,mahdavi2021software}, feature toggles have 5 types: 
\begin{enumerate*}[label=(\arabic*)]
    \item \emph{release toggle} to enable trunk-based development,
    \item \emph{experiment toggle} to evaluate new features,
    \item \emph{ops toggle} to control operational aspects,
    \item \emph{permission toggle} to provide the appropriate functionality to a user, such as special features for premium users, and
    \item \emph{development toggle} to turn on or off developmental features to test and debug.
\end{enumerate*}

Experiment toggles and permission toggles can be considered as \fsl{Domain} configurations. Ops toggles can be a kind of \fsl{Development} configuration. But none of the current values cover release toggles and development toggles. We also observed the need for a new value for release toggles while we analyzed feature toggle management systems' documentation. Hence, we define two new values for Type:
\begin{enumerate*}
    \item \fsl{Release} to cover release toggles as configurations that hide the incomplete implementation of a function from the user.
    \item \fsl{Debug} to cover development toggles that use to test and debug code.
\end{enumerate*}

\noindent\framebox[\linewidth]{
  \parbox{0.97\linewidth}{
\textbf{Type (S1, S3)}
is what is configured in the system.
Values 
are: Domain (S1, S3), Technical (S1, S3), Infrastructure (S1, S3), Development (S1), Release (S3), Debug (S3).
}}

\subsection{Artifact (S1, S2, S3)}

\fsl{Definition.} Configuration Artifacts are development artifacts that contain the configuration \cite{siegmund2020dimensions}. 

\fsl{Model Extension.} Based on Siegmund \etal's analysis the 12 artifact values are: Source Code, Configuration Code, Configuration File, Database, Command line parameter, Environment Variable, Preprocessor Code, Directory Service, Feature Model, Spreadsheet, Product Map, and GUI.  \fsl{Configuration Code} is a more complex artifact such as build scripts, container descriptions (such as Docker files), and automation scripts (such as Ansible playbook).
\fsl{Configuration File} is a file, such as properties files and .ini files. 
Meinicke {\etal} has three themes named \fsl{documentation}, \fsl{constraints}, and \fsl{dependencies} related to the Artifact dimension. They found that configuration options are documented in \fsl{Feature Models}, because configuration options have constraints, and are highly dependent on each other. However, their interviewees emphasize that feature toggles have few or no constraints, dependencies between feature toggles are nesting or grouping, and configuration knowledge is spread across \fsl{Source Code}, and \fsl{Configuration Files}.

In our feature toggle management system analysis, we did not observed any of the following artifacts: \fsl{Configuration Code}, \fsl{Preprocessor Code}, \fsl{Directory Service}, \fsl{Feature Model}, \fsl{Spreadsheet}, and \fsl{Product Map}. As we mentioned, some of them are specific to configuration options such as \fsl{Feature Model}. However, some of them can be a kind of feature toggle artifact such as \fsl{Spreadsheet}. For example, Chrome developers use a spreadsheet to record the details about toggles~\cite{chromiumsheet}. 

\fsl{Model Completion.} In the analyzed publications, feature toggle management systems are an artifact used by practitioners to implement, and manage feature toggles in their software code to control the added complexity and technical debt of adding feature toggles \cite{mahdavi2021software, ramanathanpiranha,meinicke2020capture}. 
We add \fsl{Management system} as a value to this dimension.  

\noindent\framebox[\linewidth]{
  \parbox{0.97\linewidth}{
\textbf{Artifacts (S1, S2, S3)}
are development artifacts that contain the configuration.
Values 
are: Source Code(S1,S2,S3), Configuration Code (S1), Configuration File (S1, S2, S3), Database (S1, S3), Command line parameter(S1,S3), Environment Variable (S1, S3), Preprocessor Code (S1), Directory Service (S1), Feature Model (S1, S2), Spreadsheet (S1), Product Map (S1), GUI (S1, S3), Management system (S3).}}

\subsection{Life Cycle (S1, S2, S3)}

\fsl{Definition.} The life cycle of a configuration is defined as its lifetime phases from creation to removal \cite{siegmund2020dimensions}.

\fsl{Model Extension.} Siegmund \etal~\cite{siegmund2020dimensions} identified six values: Create, Maintain, Bind, Own, Deprecate, and Remove. 

Meinicke \etal~\cite{meinicke2019exploring} identified \fsl{removal of configurations} as one of the themes. They mentioned that removal is more important and more frequent when practitioners use feature toggles because of their short-term usage goals. Configuration options tend to be permanent in the code. Meinicke~\etal also emphasized the importance of documenting the \fsl{owner} of the feature toggles in the documentation. Testing software configurations is one of the themes in Meinicke~\etal. They stated that testing is an important task for both configuration options and feature toggles. However, the strategies may differ. Configuration options have interaction with each other so the testing of their combinations should be done, but feature toggles have rare interactions, and testing them separately is enough based on practitioners' experiences. Hence, \fsl{test} should be a new value for life cycle dimension.

Analyzing feature toggle management systems documentation, we observed all the life cycle values including the new value of \fsl{test}. \fsl{Own}, \fsl{Deprecate}, and \fsl{Remove} are less frequent, and none of the repositories talk about all life cycle values.

\fsl{Model Completion.} Mahdavi-Hezaveh~\etal~\cite{mahdavi2021software} provided a life cycle for feature toggle that contains Decision, Design, Implementation, Existence, and Clean-up. Siegmund~\etal~mentioned that the life cycle of a configuration option starts when a new optional feature is planned \cite{siegmund2020dimensions}. The decision, design, and implementation steps in the feature toggle life cycle are covered by the \fsl{Create} value for the life cycle dimension. Existence is covered by \fsl{Bind}, \fsl{Maintain} and \fsl{Own}. The clean-up step in feature toggle life cycle is covered by \fsl{Deprecate}, and \fsl{Remove}. So, the life cycle of feature toggles is covered by the values for this dimension identified by Siegmund {\etal}. In MSCv2 we split \fsl{Create} to three values: \fsl{Decision}, \fsl{Design}, and \fsl{Implementation} to emphasize the importance of decision making and designing to control the added complexity to the code, and to plan removing the toggles ahead of the time~\cite{mahdavi2021software}.

 Researchers \cite{rahman2016feature,mahdavi2021software} 
mentioned that one way to remove feature toggles from code is to make the toggle a permanent configuration option in the code. So, not all the feature toggles have \fsl{Deprecate}, and \fsl{Remove} in their life cycle, and feature toggles can turn into configuration options at the end of their lifetime. Another observation from publications is the importance of testing feature toggles~\cite{rahman2016feature}. 

\noindent\framebox[\linewidth]{
  \parbox{0.97\linewidth}{
\textbf{Life Cycle (S1, S2, S3)}
of a configuration is its lifetime phases from creation to removal. Values 
are: Decision (S1, S3), Design (S1, S3), Implementation (S1, S3), Test (S2, S3), Maintain (S1, S3), Bind (S1, S3), Own (S1, S2, S3), Deprecate (S1, S2, S3), Remove (S1, S2, S3).}}

\begin{figure*}[!t]
\centering
\includegraphics[width=2\columnwidth]{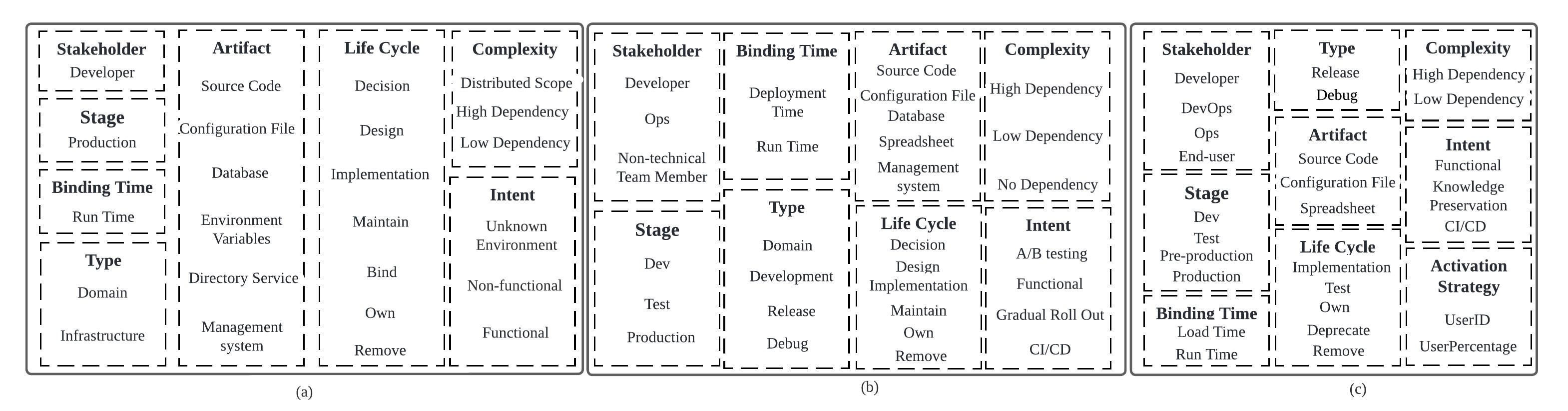}
\caption{Application of the MSCv2 on three publications on practitioners' practices: (a) \cite{sayagh2018software}, (b) \cite{mahdavi2021software}, (c) \cite{rahman2016feature}}
\label{fig:threepapers}
\end{figure*}

\subsection{Complexity (S1, S2, S3)}

\fsl{Definition.} Siegmund \etal~\cite{siegmund2020dimensions} considers two aspects to define the complexity: scope and dependency. The scope depends on the impact of the configuration on one or more modules. The dependency is the degree of dependency of a configuration to other configurations.

\fsl{Model Extension.} Siegmund~\etal~identified two possible values for scope: \fsl{Local scope} and \fsl{Distributed scope}. For dependency, they identified three possible values: \fsl{High dependency}, \fsl{Low dependency}, and \fsl{No dependency}. From the themes in Meinicke {\etal}'s \cite{meinicke2019exploring} study, \fsl{complexity}, \fsl{dependencies}, \fsl{feature traceability}, and \fsl{feature interactions} map to this dimension. They reported that practitioners try to keep the scope of feature toggle's impact as \fsl{Local Scope} in single modules. The interviewees also stated that dependencies among feature toggles are rare and interactions between them are not important, but configuration options have many dependencies often in hierarchical groups. They also found that the complexity of using configuration options is high. However, for feature toggles, the complexity depends on the number of toggles.

From the feature toggle management system documentation, we have observed \fsl{Local} (in all) and \fsl{Distributed} \cite{unleash} scopes. The practitioners did not discuss the dependency between feature toggles because the dependency is introduced while using them in the code. 
All levels of dependency are possible.

\fsl{Model Completion.} We observed all values of this dimension in analyzed publications. 

\noindent\framebox[\linewidth]{
  \parbox{0.97\linewidth}{
\textbf{Complexity (S1, S2, S3)}
is the scope and dependency between software configurations.
Values 
are: Local scope (S1, S2, S3), Distributed scope(S1,S3), High dependency(S1,S2), Low dependency (S1, S2), and No dependency (S1, S2).
}}

\subsection{Intent (S1, S2, S3)}

\fsl{Definition.} Siegmud \etal~\cite{siegmund2020dimensions} defined \fsl{intent} as the purpose of using software configuration. Meinicke \etal~\cite{meinicke2019exploring} identified \fsl{goal} as a theme that is defined same as \fsl{intent}. 


\fsl{Model Extension.} From our analysis of documentations, we observed that feature toggles are used for A/B testing \cite{qandidatetoggle}, gradual rollout (canary release) \cite{flipper}, turning on and off a feature \cite{unleash}, targeted release \cite{unleash}, implementing trunk-based development (prevent long-lived branches) \cite{flipper}, release new feature faster \cite{flipper}, and as kill switches~\cite{flipper}. 

Siegmund {\etal} identified the following values fot \textit{Intent} dimension: A/B Testing, Code Reuse, Knowledge Preservation, Reduced Testing Effort, Distributed Environment, Unknown Environment, Non-Functional, and Functional. From these identified values, \fsl{A/B testing} can cover A/B testing for feature toggles. \fsl{Functional} value can cover turning on and off a feature, releasing new features faster, and use as kill switches from feature toggles usage purposes. The rest of the intentions of feature toggle usages that we found in our analysis are not covered by the identified values by Siegmund~\etal Meinicke {\etal} distinguish three main goals: hiding incomplete implementation, experimentation and release, and configuration. Based on these goals' definitions, \fsl{hiding incomplete implementation} can cover implementing trunk-based development, \fsl{experimentation and release} can cover A/B testing and gradual rollout, and \fsl{configuration} is same as \fsl{functional} value in Siegmund's values. Also, Meinicke~\etal~stated that the goal behind a feature toggle can be changed over time. For example, an experiment feature toggle with the goal of A/B testing can become a configuration option with the goal of changing functionality based on user request. Other researchers also mentioned this situation in their studies~\cite{rahman2016feature, mahdavi2021software}.

\fsl{Model Completion.} We observed that feature toggles are also used for CD, context switching, migration from one environment to another, and blue/green deployment in analyzed publications \cite{rahman2016feature,hoyos2021removal}. Context switching, when a developer is working on a feature and an emergency bug fixes came up for another feature, is covered by \fsl{knowledge preservation}. Migrating from one environment to another environment and blue/green deployment are covered by \fsl{Unknown Environment}. 

Based on new observed values in the feature toggle management systems' documents, goals in \cite{meinicke2019exploring}, and analyzed publications, we define the following new values for this dimension:
\begin{enumerate*}
    \item \fsl{Gradual Roll Out}: when a feature will be exposed to a subset of users and gradually is made available for all users. This value covers gradual rollout and targeted releases in the list of feature toggle usage intents,
    \item \fsl{CI/CD}: when using toggles helps to implement trunk-based development and prevents having long-lived branches. Moreover, the development team can release the system rapidly even with long-term feature development and help them to do dark launches.
\end{enumerate*}

\noindent\framebox[\linewidth]{
  \parbox{0.97\linewidth}{
\textbf{Intent (S1, S2, S3)}
is the purpose of using software configuration. 
Values 
are:  A/B Testing (S1, S2, S3), Code Reuse (S1), Knowledge Preservation (S1, S3), Reduced Testing Effort (S1), Distributed Environment (S1), Unknown Environment (S1, S3), Non-Functional (S1), Functional (S1, S2, S3), Gradual Roll Out (S2, S3), CI/CD (S2, S3).}}

\subsection{Activation Strategy (S3)}

\fsl{Definition.} 
We define \fsl{Activation Strategy} as rules to enable a  feature toggle for specific users or a subset of users.

\fsl{Model Extension.} While we analyzed feature toggle management system documents, we identified a new dimension specifically related to feature toggles: \fsl{Activation Strategy}. 

\fsl{Model Completion.} The values we have identified \fsl{UserID}~\cite{rahman2016feature,hoyos2021removal}, \fsl{GroupID}~\cite{hoyos2021removal}, \fsl{UserPercentage}~\cite{rahman2016feature,ramanathanpiranha}, and \fsl{IPs}~\cite{ramanathanpiranha,hoyos2021removal} are observed in the publications as well.

We observed the following activation strategy values in our analysis:
\begin{enumerate*}
    \item \fsl{Default}: the feature is enabled for all users \cite{flipper};
    \item \fsl{AlwaysOn}: the feature is always enabled for all users \cite{FeatureToggle};
    \item \fsl{AlwaysOff}: the feature is always disabled for all users \cite{FeatureToggle},
    \item \fsl{UserID}: the feature is enabled for users with specified user ids. Users can select randomly or specifically \cite{flipper};
    \item \fsl{GroupID}: the feature is enabled for users in groups with specified group id \cite{unleash};
    \item \fsl{IPs}: the feature is enabled for a list of IPs. Using IP information, the feature could be enabled for users in a particular region or country as well \cite{unleash};
    \item \fsl{HostNames}: the feature is enabled for a list of hostnames \cite{unleash};
    \item \fsl{UserPercentage}: the feature is enabled to a percentage of users. In gradual rollout when the percentage increases, previous users should remain in \cite{flipper};
    \item \fsl{TimePercentage}: the feature is enabled for a percentage of time \cite{flip};
    \item \fsl{TimePeriod}: the feature is enabled on or after a date, on or before a date, between two dates, or on specified days of the week \cite{FeatureToggle};
    \item \fsl{Random}: the feature is enabled or disabled randomly \cite{FeatureToggle};
    \item \fsl{Computed}: the feature is enabled based on the values of other feature toggles \cite{flag};
    \item \fsl{Flexible}: which combines more than one strategy into one strategy to have a complex strategy. An example is combining the UserID strategy, and Percentage strategy \cite{unleash}; and 
    \item \fsl{Custom}: in some cases, developers need a specific strategy that is not implemented as a built-in activation strategy. Some feature toggle packages allow developers to define a custom activation strategy~\cite{flags}.
\end{enumerate*}

\noindent\framebox[\linewidth]{
  \parbox{0.97\linewidth}{
\textbf{Activation Strategies (S3)}
are rules to enable a feature toggle for specific users or a subset of users.
Values 
are: Default (S3), AlwaysOn (S3), AlwaysOff (S3), UserID (S3), GroupID (S3), IPs (S3), HostNames (S3), UserPercentage (S3), TimePercentage (S3), TimePeriod (S3), Random (S3), Computed (S3), Flexible (S3), Custom (S3).
}}

\begin{figure}[!t]
\centering
\includegraphics[width=0.9\columnwidth]{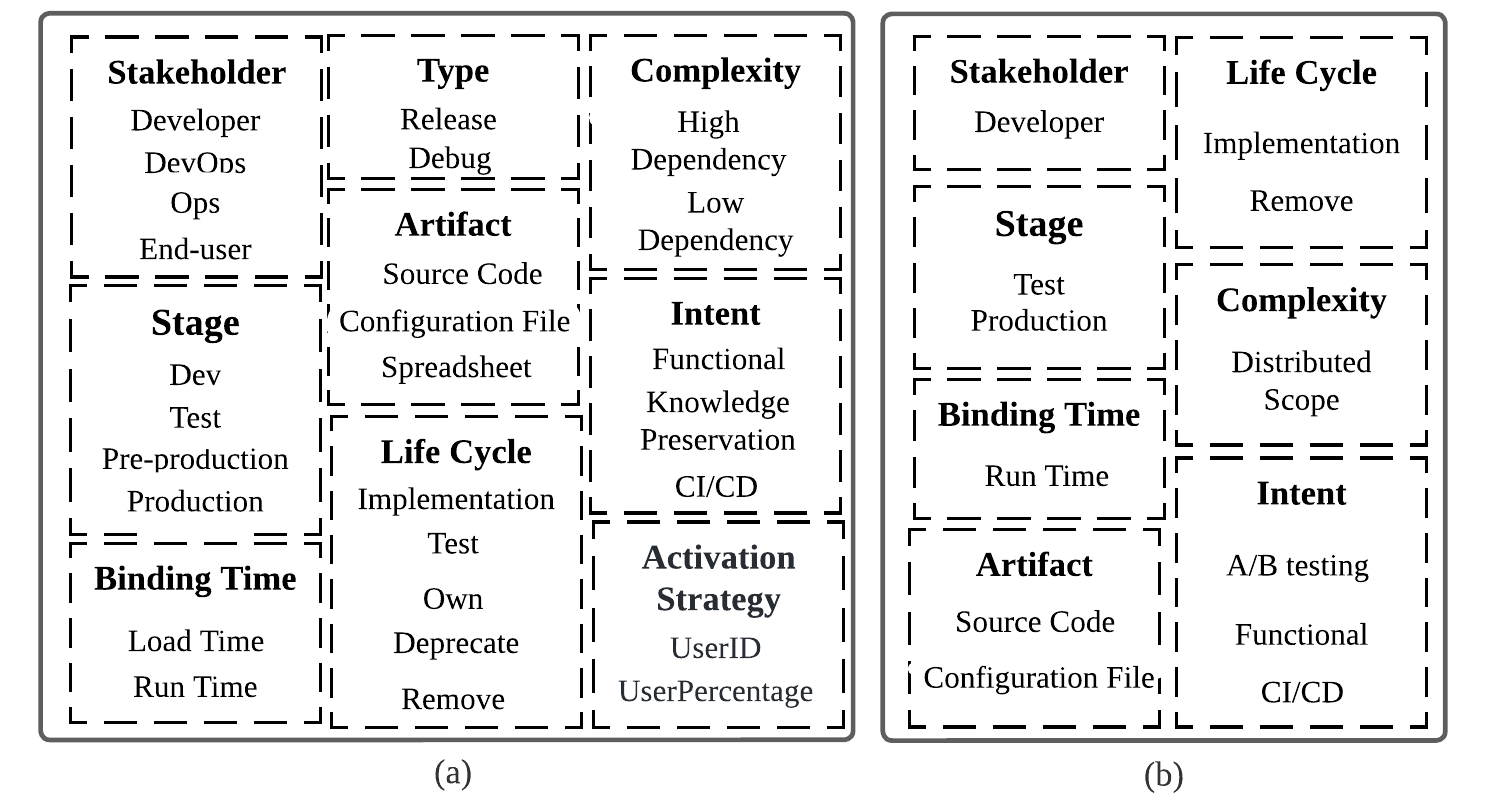}
\caption{Application of the MCSv2
: (a) \cite{rahman2016feature}, (b) \cite{rahman2018modular}}
\label{fig:papercontext}
\end{figure}

\section{Application of MSCv2}
\label{discussion}

In this section, we show how MSCv2 can be used by researchers and practitioners (\textbf{answer to RQ2}). 
As explained in Subsection~\ref{method_phase3}, to use MSCv2, researchers and practitioners should select appropriate values for each dimension to describe software configuration in their system. The result will be an instance of MSCv2. 
In Subsection~\ref{subsec:sameRQinthree}, we present analysis of three publications that ask  similar research questions about feature toggles and configuration options.
In Subsection~\ref{subsec:chromeintwo}, we discuss two publications that perform research on Chrome. 
In Subsection~\ref{subsec:fiveconfiginchrome}, we show how the MSCv2 can be used to model existing  configurations in the Chrome repository.

\begin{figure*}[!t]
\centering
\includegraphics[width=1.3\columnwidth]{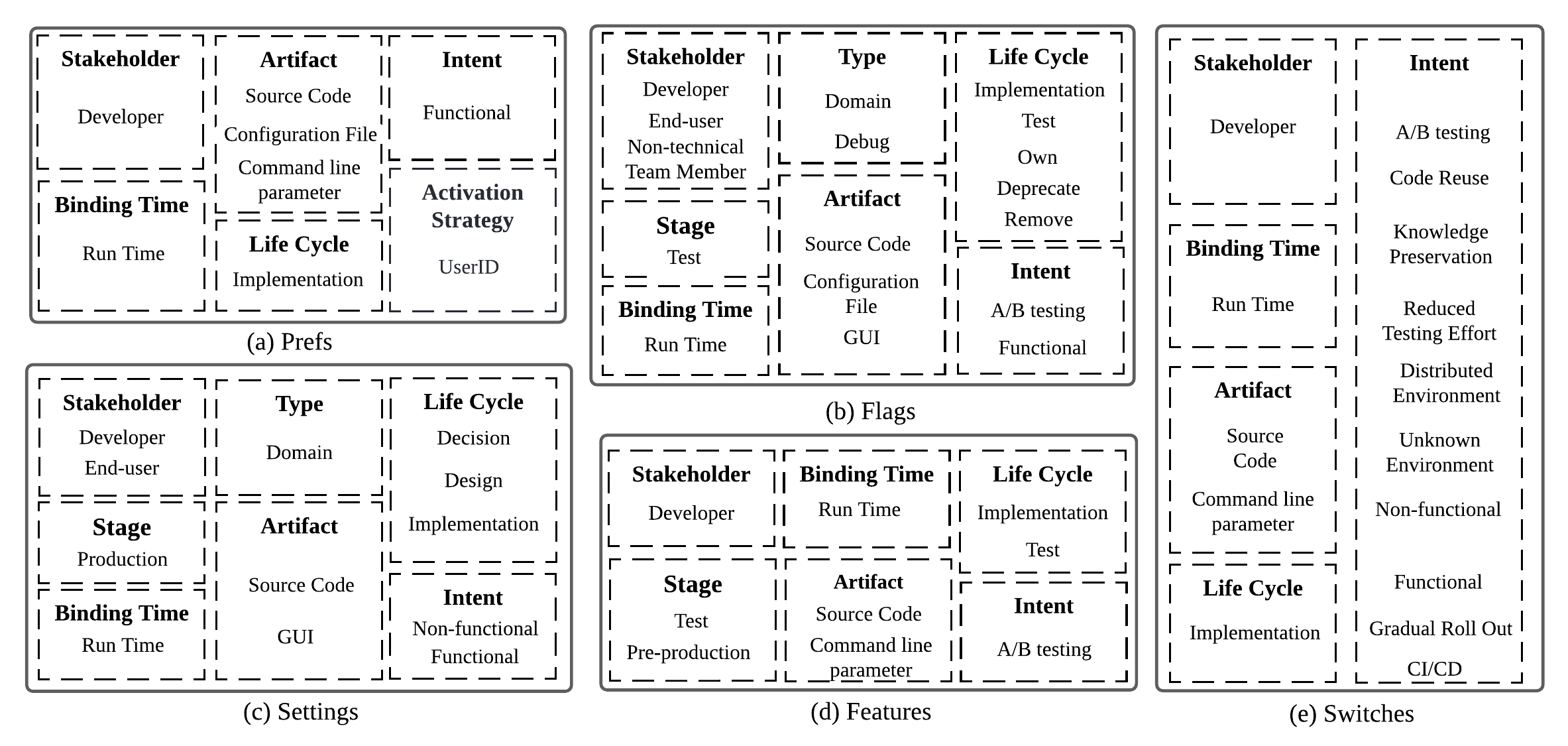}
\caption{Application of the MSCv2 on Chrome}
\label{fig:chromecontext}
\end{figure*}

\subsection{Similar Papers on Industry Practices}
\label{subsec:sameRQinthree}

We selected three publications with similar research question of practices used by practitioners to 
use software configuration. 
Rahman {\etal} performed a thematic analysis of 17 videos and blog posts from practitioners to understand the challenges, and advantages of using \textit{feature toggles} \cite{rahman2016feature}. Mahdavi-Hezaveh {\etal} analyzed 109 gray literature artifacts and publications to identify \textit{feature toggle} usage practices~\cite{mahdavi2021software}. Sayagh {\etal} interviewed 14 practitioners, surveyed 229 practitioners, performed a systematic literature review of 106 publications, and identified activities, challenges, and recommendations related to \textit{run-time configuration options}~\cite{sayagh2018software}.

To derive models for these three publications, we followed the steps discussed in Section~\ref{method_phase3}, and the result is visualized in Figure~\ref{fig:threepapers}. 
The publications may indicate that the research areas of these studies are different. However, looking at their MSCv2 instances in Figure~\ref{fig:threepapers} the studies have numerous similarities. Values for all dimensions have overlapped in these three instances.  
Considering the similarities and differences, the results in these studies can be compared or integrated. 
Mahdavi-Hezaveh~\etal \cite{mahdavi2021software} discussed the partial overlap of seven out of 17 practices they identified with Sayagh~\etal \cite{sayagh2018software} study results. 
Mahdavi-Hezaveh~{\etal} also used Rahman~\etal~\cite{rahman2016feature} as one of the analyzed artifacts for their study. 

New researchers in the software configuration research area can easily miss related publications if they are not aware of using different keywords of \textit{feature toggles} and \textit{configuration options} by other researchers. Also, if researchers 
populate an instance of MSCv2 in their publications
other researchers can use their results more easily to develop new studies.

\subsection{Studies on Chrome repository}
\label{subsec:chromeintwo}

We modeled the software configurations of Chrome from two publications \cite{rahman2016feature, rahman2018modular} as shown in Figure~\ref{fig:papercontext} using MSCv2. In \cite{rahman2016feature}, Rahman~\etal~ quantitatively analyzed the usage of feature toggles in 39 releases of Chrome. In \cite{rahman2018modular}, the authors extracted four architectural representations of Chrome including feature toggle architecture. The same software repository is used in both publications. While modeling Chrome in these publications, our results indicate that even when the researchers analyze the same system, they may not define the configuration the same due to the lack of a comprehensive model for defining software configuration. Some values for dimensions exist in both model instances but some dimensions and some values exist in just one of the model instances. If researchers looked at two models in Figure~\ref{fig:papercontext} without having knowledge about these two publications, they may claim that these two systems have similarities but they are not the same.

Researchers can use MSCv2 when designing their studies, and show the characteristics of software configuration in a MSCv2 instance. Looking at the instance, other researchers will have the same viewpoint of the system for their analysis. 

\subsection{Chrome Repository}
\label{subsec:fiveconfiginchrome}

To better understand the configurations in the Chrome, we applied the MSCv2 on its repository documentation (as we discussed in Section~\ref{method_phase3}) to find values for each dimension for each configuration. The result is shown in Figure~\ref{fig:chromecontext}. In the Chrome documentation, five different terms are used for the software configuration: 1) Prefs, 2) Flags, 3) Settings, 4) Features, and 5) Switches. Based on the names of these configurations, the difference between them is unclear. For instance, in the research literature and gray literature switches and flags are the same concepts \cite{rahman2016feature,fowlerfeature}, but in the Chrome repository, these terms seem to be two separate concepts. Figure~\ref{fig:chromecontext} indicates similarities and differences between these five MSCv2 instances. For example, developers are stakeholders of all five configurations, but non-technical team members only have access to Flags. We did not find values for some of the dimensions. For example, the documentation does not include information related to the \fsl{Complexity} dimension for any of the configuration terms. Architectures can use MSCv2 to define configuration in software projects to make sure that every characteristic of a configuration is clear to team members. If developers do not have a correct understanding of the concept behind each configuration, they may use them incorrectly. 

We can compare the models in Figure~\ref{fig:papercontext} and Figure~\ref{fig:chromecontext}. In the two publications \cite{rahman2016feature, rahman2018modular} that are shown in Figure~\ref{fig:papercontext}, Rahman {\etal} consider \textit{Switches} in the Chrome repository as feature toggles in the system. 
Hence, both models in Figure~\ref{fig:papercontext} should be the same as Switches (e) in Figure~\ref{fig:chromecontext}. We observe that not only these three models are not the same but also they are not even subsets of each other. This observation shows that using unstructured documentation to define configuration in industry and in research publications may result in an incomplete and inconsistent understanding of a configuration definition. 



\subsection{Takeaways}

The takeaways are based on 
Sections~\ref{subsec:sameRQinthree}--~\ref{subsec:fiveconfiginchrome}.

\textbf{For researchers.} Researchers may define software configuration differently in their publications even if the industrial system is the same (Sections~\ref{subsec:chromeintwo} and \ref{subsec:fiveconfiginchrome}). They can use MSCv2 by creating an instance and selecting proper values for each dimension to model the software configuration in the system of their publications. Additionally, by using MSCv2 researchers can compare and integrate their results with similar studies, prevent duplication of effort, increase the impact of their research result through a bigger family of research, and meta-analyze the results of a family of research publications~(Section~\ref{subsec:sameRQinthree}). Also, using MSCv2 to model the configuration of industry systems by software architects when designing configurations, could help researchers to record the configuration more similar to what might be done by a practitioner, and enable knowledge transfer between industry and academia~(Sections\ref{subsec:chromeintwo}).

\textbf{For practitioners.} 
An industrial system can have more than one kind of software configuration (Section~\ref{subsec:fiveconfiginchrome}), so a systematic definition of configurations by using MSCv2 can make the definitions more coherent and precise.

\section{Threats to Validity}
\label{limitations}

In Phase~1 of the methodology, we used keyword search 
to find related repositories. We may have missed related repositories if the keywords are not used in the name and description of the repository. In addition, in Step 4 of the filtering process of the search results, we read the name and description of the repositories. We did not check the content of the repositories. 
However, since we selected the top 20 repositories in our result and checked the content of them, this filtering process did not affect our results. We selected 20 repositories from the top 5 programming languages and analyzed their documentation. The analysis of more repositories may enrich the results of this research study. To overcome this limitation, we defined the stopping criteria for analyzing documentation when we did not find any new dimension and any new value.

In Phase~2 of the methodology, we analyzed the feature toggle management systems' documentation using coding practices. To prevent subjective results, the coding was done by the first and the second authors separately. Then we discussed and resolved disagreements. We may have missed some values because practitioners may use feature toggle in a different way that is not mentioned in the documentation. 

 In Phase~3 of the methodology, we applied MSCv2 to five publications and Chrome. To show the generalization of MSCv2, the publications were selected to be related to either feature toggles or configuration options, and the process was done by two individuals. In future work, we will ask other researchers to apply the MSCv2 in software configuration studies and share their observations.

\section{Conclusion and Future Work}
\label{conclusion}

Feature toggles and configuration options are used widely in the software development process for different purposes, such as practicing CI/CD and system customization. Feature toggles and configuration options fall under the same umbrella with different names. 
Hence, we extended the existing model of software configuration (MSC) to cover both feature toggle and configuration option concepts. The MSCv2 has 9 dimensions with 70 values. 
We observed that MSCv2 may help researchers to build a comprehensive model to define the configuration of analyzed systems in their research studies. Using MSCv2 may also help practitioners to define their system's configuration, and provide a common language for two communities to transfer knowledge about software configurations.

A future path for this study is to compare the configuration definition in feature toggle-related publications and configuration option-related publications using MSCv2. The results can lead researchers to find new research opportunities, transfer knowledge between two research areas, and better understand this family of research. Also, the researchers can analyze relationships between the dimensions and values.  


\bibliographystyle{IEEEtran}
\bibliography{dimension}


\end{document}